\documentclass[reprint,floatfix,amsmath,amssymb,aps]{revtex4-2}

\usepackage{graphicx}
\usepackage{dcolumn}
\usepackage{bm}
\usepackage{bbold}
\usepackage{float}
\usepackage{braket}
\usepackage{dsfont}
\usepackage{color}
\usepackage{float}
\usepackage{xcite}
\usepackage{xr-hyper}
\usepackage{hyperref}
\makeatletter
\newcommand*{\addFileDependency}[1]{
  \typeout{(#1)}
  \@addtofilelist{#1}
  \IfFileExists{#1}{}{\typeout{No file #1.}}
}
\makeatother

\newcommand*{\myexternaldocument}[1]{%
    \externaldocument{#1}%
    \addFileDependency{#1.tex}%
    \addFileDependency{#1.aux}%
}

\myexternaldocument{supplement}

\begin{document}

\preprint{APS/123-QED}

\title{
 Lyapunov spectrum scaling  for classical many-body dynamics close to integrability
}

\author{Merab Malishava$^{1,2}$}
\email[Corresponding author:\\]{merabmalishava@gmail.com}
\author{Sergej Flach$^{1,2}$}
\email[Corresponding author:\\]{sergejflach@googlemail.com}
\affiliation{%
\mbox{$^1$Center for Theoretical Physics of Complex Systems, Institute for Basic Science(IBS), Daejeon, Korea, 34126}\\
$^2$Basic Science Program, Korea University of Science and Technology(UST), Daejeon, Korea, 34113
}

\date{\today}
\begin{abstract}
We propose a novel framework to characterize the thermalization of many-body dynamical systems close to integrable limits using the scaling properties of the full Lyapunov spectrum. We use a classical unitary map model to investigate macroscopic weakly nonintegrable dynamics beyond the limits set by the KAM regime. We perform our analysis in two fundamentally distinct long-range and short-range integrable limits which stem from the type of nonintegrable perturbations. Long-range limits result in a single parameter scaling of the Lyapunov spectrum, with the inverse largest Lyapunov exponent being the only diverging control parameter and the rescaled spectrum approaching an analytical function. Short-range limits result in a dramatic slowing down of thermalization which manifests through the rescaled Lyapunov spectrum approaching a non-analytic function. An additional diverging length scale controls the exponential suppression of all Lyapunov exponents relative to the largest one. 
\end{abstract}

\maketitle

Thermalization is a universal property of the long-time dynamics of generic nonintegrable many-body systems. Thermal equilibrium is characterized by stationary distributions and assumes ergodicity and mixing in the phase space \cite{huang1987statistical}. The thermalization dynamics will in general slow down close to integrability, and may even
cease to be observed  \cite{gogolin2011absence,rigol2009breakdown, campbell2005introduction, gaveau2015ergodicity, bouchaud1992weak, bel2006weak, bel2005weak}, which was also noted in earlier studies of dynamical systems \cite{ford1992fermi, zabusky1967dynamics,todaVibration1967, henon1964applicability}. The theory of weak nonintegrable perturbations for finite Hamiltonian systems was pioneered by Kolmogorov in 1954 \cite{kolmogorov1954conservation} and later by Arnold \cite{arnold2009proof} and Moser \cite{moser1962invariant}.
The corresponding KAM theory demonstrates the violation of the ergodic hypothesis for sufficiently weak perturbations due to the emergence of a mixed phase space with a finite fraction of points belonging to regular trajectories on tori.
At a critical strength of the nonintegrable perturbation, all tori disappear, and the dynamics become fully chaotic allowing for thermalization.
But what exactly is meant by ``sufficiently weak" and ``critical strength"? As it turns out the magnitude of the critical perturbation $\tilde H$ decays rapidly with the growth of the number of degrees of freedom. Namely, $|\tilde H|\leq a N^{-b}$ with $b = 160$ has been shown as an upper bound for applicability of KAM theory in lattice systems with short-range interactions (for example an array of Josephson junctions) \cite{wayne1984kam}. This result suggests that it is practically impossible to witness quasiperiodic motion suggested by KAM in macroscopically large systems
close to integrability. What is then the expected behavior of systems with a large number of degrees of freedom in proximity to an integrable limit? How does one characterize it? Does it have universality classes? Is there a KAM-like regime for macroscopic models? If not, what lies beyond the KAM horizon spanned by finite systems?
\begin{figure}
    \centering
    \includegraphics[width=\linewidth]{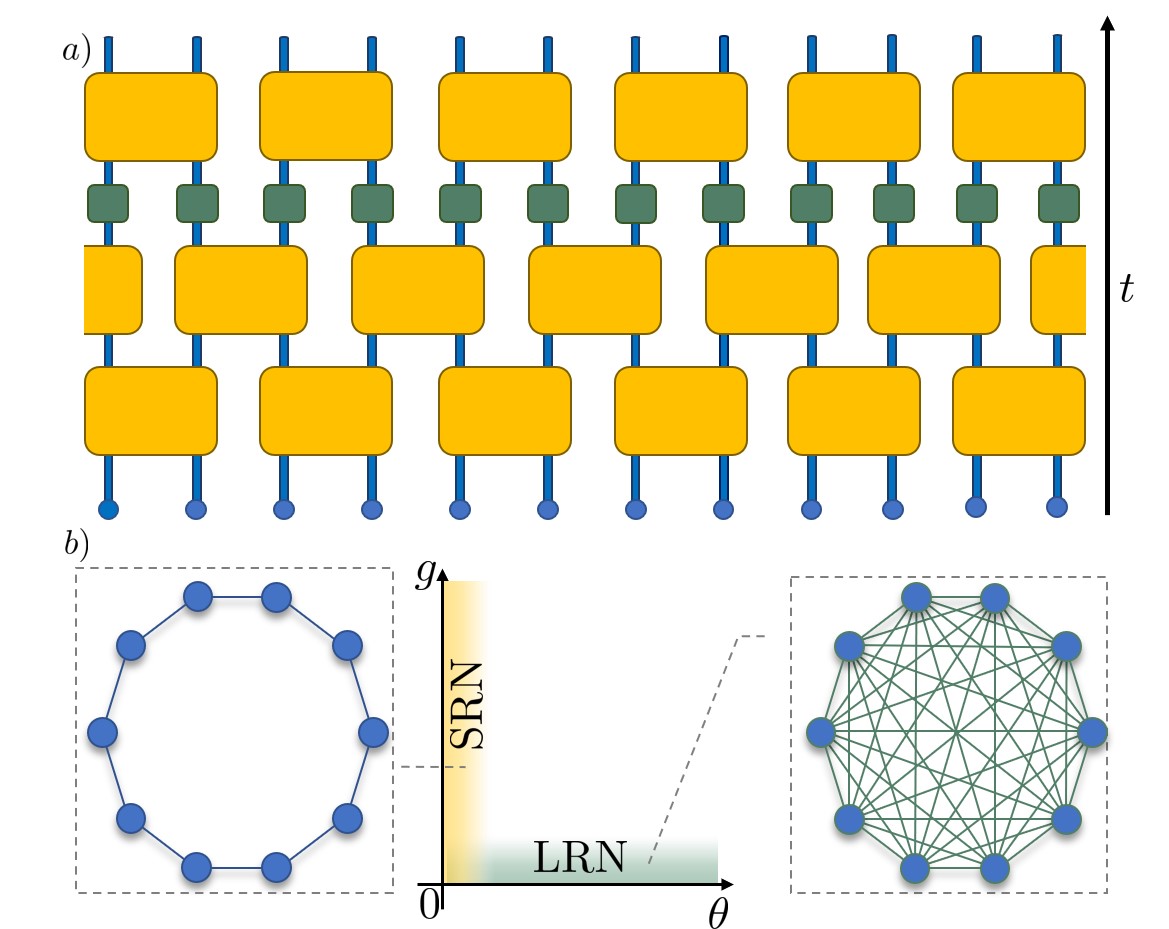}
    \caption{a) A schematic representation of the unitary circuits map. Large yellow blocks indicate $\hat C$ unitary matrices parametrized by the angle $\theta$. Small green blocks indicate local nonlinearity generating map $\hat G$ parametrized by the nonlinearity strength $g$. The black arrow on the right indicates the time flow. b) Control parameter space $\{\theta,g\}$ with the highlighted area corresponding to the induced networks. Integrable limits are reached for $g=0$ (linear evolution of extended normal modes) or $\theta=0$ (decoupled nonlinear map). Small nonzero $g$ values induce LRNs, small nonzero $\theta$ values induce SRNs. The network images indicate actions (filled circles) and couplings induced by the nonintegrable perturbation (straight lines). Left image - SRN, right image - LRN.
}
    \label{fig1}
\end{figure}

\begin{figure}
    \centering
    \includegraphics[width=\linewidth]{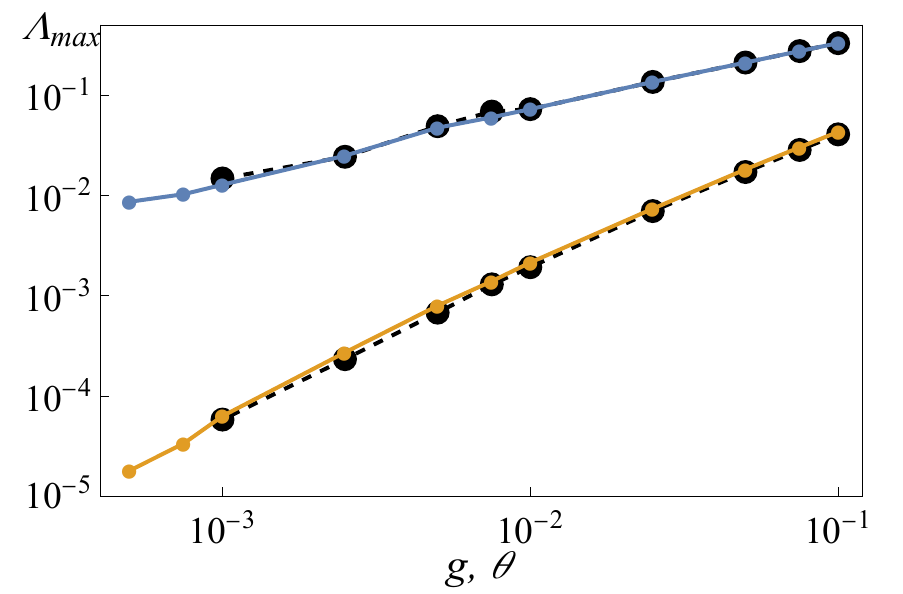}
    \caption{The largest Lyapunov exponents $\Lambda_\text{max}$ in SRN (blue small circles, top) and LRN (orange small circles, bottom) regime versus the corresponding deviation from integrable limit $g, \theta$. Solid lines connect the data points and guide the eye.
    For the SRN case, the parameter nonlinearity is fixed $g = 1.0$, while for the LRN case the angle $\theta$ is fixed at $0.33\pi$. For both cases system size $N = 200$. The large black circles connected by dashed lines correspond
    to data for system size $N = 100$.}
    \label{fig2}
\end{figure}

To quantify the thermalization of a system one typically chooses a specific set of observables and studies
equipartition and ergodicity for those specific observables. In their pioneering work Fermi, Pasta, Ulam, and Tsingou attempted to showcase equipartition using the normal modes of a linear chain as their choice of observables for a weakly nonlinear chain \cite{fermi1955studies}. In the absence of nonlinearity, the normal modes are ``frozen", i.e. they become the \textit{actions} of the integrable system. 

Recent studies attempted to broaden the ergodicity analysis by computing convergence of finite-time average distributions of observables to their phase space averages. They revealed that most physical systems belong to two distinct classes when it comes to thermalization in proximity to integrable limits \cite{danieli2019dynamical, mithun2019dynamical, thudiyangal2021fragile}. Systems with weak nonlinear perturbations such as FPUT chains, Josephson junction networks in the limit of small energy density, discrete nonlinear Schr\"odinger equations all belong to the class of \textit{Long Range Networks} (LRN). On the other hand, a broad range of lattice systems allowing for proximity to an integrable limit 
of vanishing lattice coupling belongs to a class of \textit{Short Range Networks} (SRN). Examples of SRN include coupled anharmonic oscillator chains in the limit of weak coupling \cite{danieli2019dynamical}, 
Josephson junction chains in the limit of weak Josephson coupling \cite{mithun2019dynamical}, etc.

There are serious limitations of studying thermalization through observable dynamics. The choice of observables is ambiguous
\cite{kurchan2019equilibration, ganapa2020thermalization}, and even for integrable systems specifically chosen observables show ergodic thermal-like behavior \cite{baldovin2021statistical}. Observable dynamics address ergodicity, but not mixing. However, nonintegrable dynamics
are necessarily mixing, show typically exponential decay of correlations with a macroscopic set of correlation times.

In this Letter, we overcome the above limitations by computing the entire Lyapunov spectrum \cite{oseledets1968multiplicative}. Lyapunov spectra
were previously used for diagnosing phase transitions \cite{fine2015chaotic} and energy localization \cite{iubini2021chaos}. Here we show that the scaling properties of the Lyapunov spectrum offer a conceptual novel way for the description of weakly nonintegrable dynamics in a generic model setup. We consider a macroscopically large system beyond the limits set by KAM and characterize thermalization in both SRN and LRN regimes, thus drawing a very general picture that encapsulates a great number of physically realizable scenarios and is directly applicable to most weakly nonintegrable classical systems. 


Resolving the entire Lyapunov spectrum for a large system is a numerically challenging task. It relies on the simultaneous evolution of a large number of trajectories, \cite{benettin1980lyapunov}. The proximity to integrable limits makes this task even harder due to an increase of the thermalization times.
In view of these challenges, we need models which possess all physically relevant features to achieve thermalization and are extremely efficient for the numerical evolution - unitary maps. The fast, exact, error-free discrete-time evolution is a key feature of unitary maps which makes them advantageous for heavy numerical tasks. These properties were on display in recent studies, where discrete unitary maps were used to achieve record-breaking evolution times for nonlinear wave-packet spreading tasks \cite{vakulchyk2019wave}, Anderson localization \cite{vakulchyk2017anderson, malishava2020floquet}, and soliton dynamics \cite{vakulchyk2018almost}.

\textit{Model --}
We use classical Unitary Circuit maps. We define a 1D lattice of size $N$ with one complex component $\psi_n$ per site $n$.
The classical dynamics evolves the vector $\vec\Psi = \lbrace \psi_n \rbrace$ in a corresponding phase space of dimension $2 N$ 
on a deterministic trajectory specified by an initial condition. 
The evolution is performed by subsequent applications of the map:
\begin{eqnarray}
\hat U = \sum_{n \in \mathds{Z}} \hat G_n \sum_{n \in 2 \mathds{Z} + 1}\hat C_{n, n+1} \sum_{n \in 2 \mathds{Z}} \hat C_{n, n+1}.
\label{evolution_operator}
\end{eqnarray}
The unitary matrices $\hat C_{n, n+1}$ 
are parametrized by the rotation angle $\theta$ and act as hoppings on pairs of neighboring sites:
\begin{eqnarray}
\hat C_{n, n+1} \begin{pmatrix}
\psi_n(t) \\ \psi_{n+1}(t)
\end{pmatrix}  = 
\begin{pmatrix}
\cos\theta & \sin\theta \\
-\sin\theta & \cos\theta
\end{pmatrix}\begin{pmatrix}
\psi_n(t) \\ \psi_{n+1}(t)
\end{pmatrix}  ,\nonumber\\
\end{eqnarray}
and the local map $\hat G_n$ induces nonlinearity:
\begin{eqnarray}
\hat G_n \psi_n = e^{i g |\psi_n|^2} \psi_n.
\end{eqnarray}
The classical unitary circuit dynamics is schematically represented in Fig. \ref{fig1}.

The map dynamics of classical unitary circuits is mixing and therefore ergodic. The map possesses two distinct LRN and
SRN integrable limits. We checked that the thermalization properties of observables in unitary circuits are in line with previous observations for Hamiltonian
systems \cite{danieli2019dynamical,mithun2019dynamical}.

\begin{figure}[t]
    \centering
    \includegraphics[width=\linewidth]{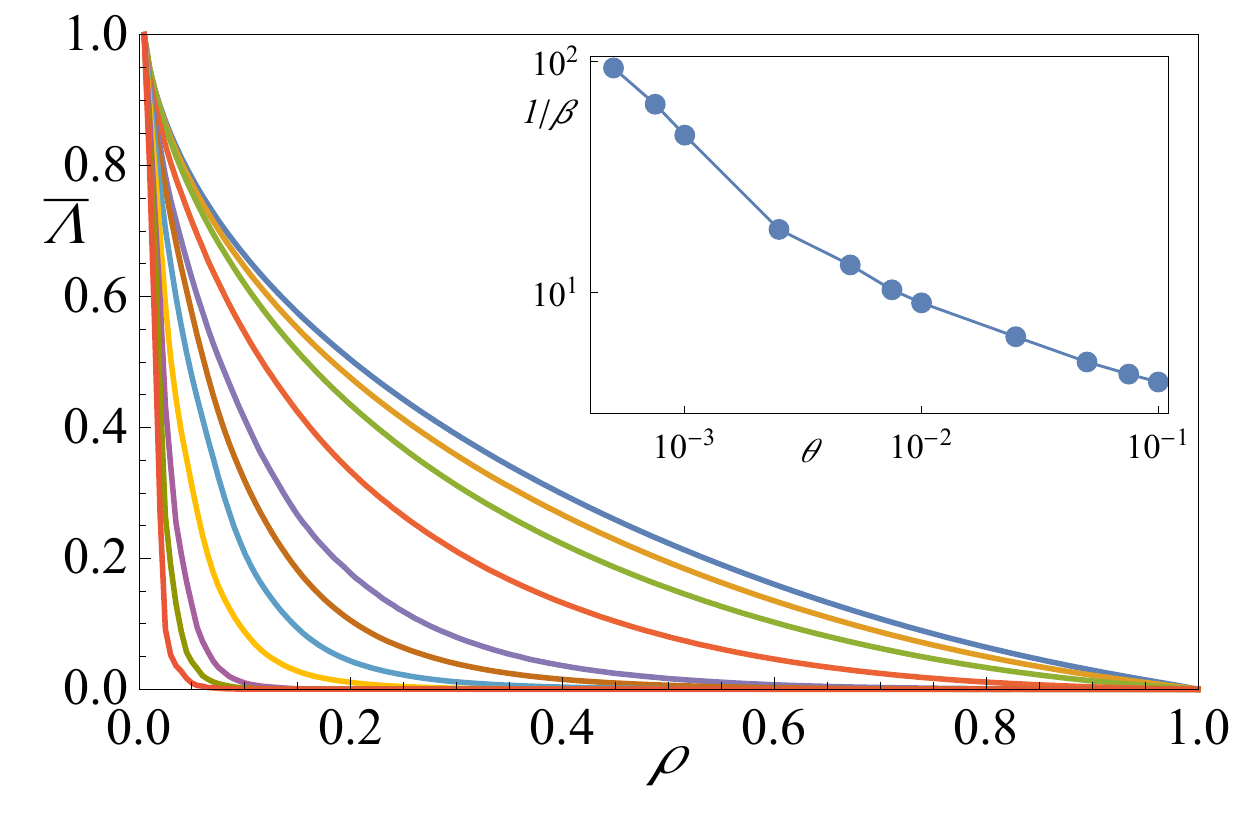}
    \caption{Renormalized Lyapunov spectrum $\lbrace \bar \Lambda_i \rbrace$ for SRN in proximity to the integrable limit. The nonlinearity strength is fixed at $g = 1.0$. Angle $\theta$ varies from $10^{-1}$ (blue, top) to $5 \cdot 10^{-4}$ (red, bottom). The inset shows the coefficient $1/\beta$ of the exponential decay (see also Fig. \ref{fig4}) of the curves as a function of $\theta$. System size $N = 200$.}
    \label{fig3}
\end{figure}

\textit{SRN integrable limit --}
We consider the limiting case $\theta = 0$ with a fixed nonzero value of nonlinearity strength $g$. In this setup the matrices $\hat C_{n, n + 1}$ become unity, thus decoupling the sites. The unitary evolution applies a nonlinear norm dependent phase shift at each site:
\begin{eqnarray}
\hat U_\text{int}^\text{SRN} \vec\Psi= \sum_{n \in \mathds{Z}} e^{i g |\psi_n|^2} \psi_n.
\end{eqnarray}

The system turns integrable with the norm at each site $|\psi_n|$ being a constant of motion. By introducing a weak deviation from the limit $0 < \theta \ll 1$ one induces a network with next-to-nearest-neighbor hopping - an SRN (see Supplemental Material for more details \cite{[{See the Supplementary Material at the [URL]}]SM}). We schematically represent the parameter space and corresponding network in Fig. \ref{fig1}(b). 

\textit{LRN integrable limit --} 
Vanishing nonlinearity strength $g = 0$ results in a linear evolution with corresponding eigenvalue problem $e^{i \omega} \vec{\Psi}(t) = \hat U_\text{int}^\text{LRN} \vec{\Psi}(t)$ (see details in Supplemental Material \cite{SM}). The evolution corresponding to this integrable limit is given by:
\begin{eqnarray}
\hat U_\text{int}^\text{LRN}\vec\Psi = \sum_k \hat u_k \vec\psi_k,
\end{eqnarray}
where $\vec{\psi}_k$ are the normal modes of the system and $u_k$ is the evolution map in reciprocal space corresponding to the wavenumber $k$. In this limit the absolute values of the normal mode amplitudes $|c_k| = |\vec\psi_k^\dagger \cdot \vec\Psi(t)|$ are the constants of motion. The deviation from this limit $g \neq 0$ induces an all-to-all coupling among the normal modes of the system which respects translational invariance through selection rules \cite{SM}. This by definition constitutes an LRN.  The green region in the control parameter space in Fig. \ref{fig1}(b) corresponds to that LRN with the schematic representation of the network sketched right to it.

\begin{figure}[!t]
    \centering
    \includegraphics[width=\linewidth]{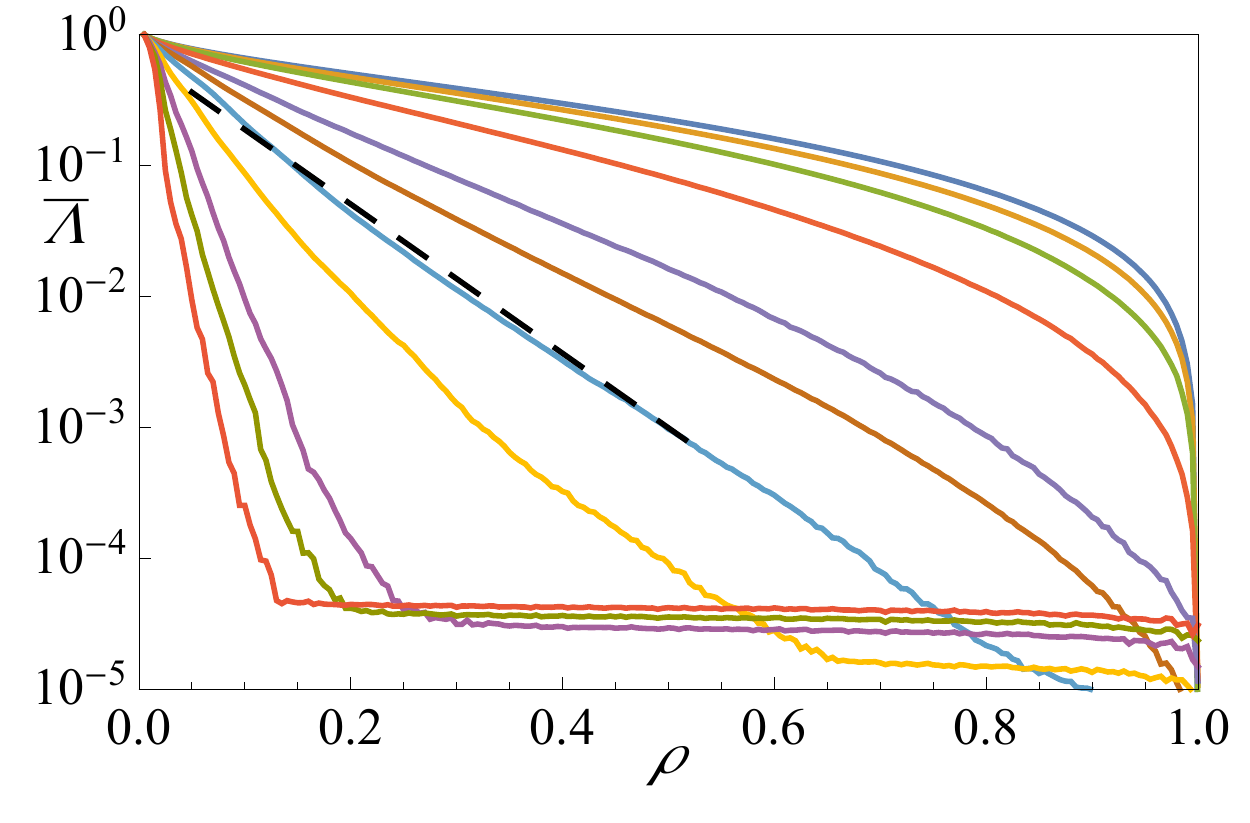}
    \caption{Renormalized Lyapunov spectrum $\lbrace \bar \Lambda_i \rbrace$ for SRN in log scale (corresponding to the parameters and data of Fig. \ref{fig3}). The dashed line is to guide the eye for the fit of exponential decay.}
    \label{fig4}
\end{figure}

We compute the Lyapunov spectrum of Unitary Circuits in proximity to integrable limits in order to resolve the entire set of characteristic time scales. We follow the evolution of a set of orthogonal tangent vectors  $\lbrace \vec w_i \rbrace$ in the $2N$ dimensional phase space and compute the increment $\gamma_i(t) = |\vec w_i(t)|$. Details on the approach can be found in Section IV of the supplemental material \cite{SM}.
For each vector we compute the transient value $X_i(t) = 1/t\sum_\tau^t \log \gamma(\tau)$ which in the infinite time limit turns into the Lyapunov characteristic exponent (LCE) $\Lambda_i = \lim_{t\rightarrow\infty}X_i(t)$. The numerically computed LCEs are the values of $X(t)$ extracted at the last step of the dynamics. 
The LCEs are ordered from largest to smallest value upon incrementing the index $i$.
Due to the symplectic nature of the map the spectrum is symmetric with LCEs coming in pairs $\Lambda_i = -\Lambda_{2N - i + 1}$. Norm conservation ensures two vanishing LCEs $\Lambda_N = \Lambda_{N+1} = 0$ \cite{skokos2010lyapunov}. According to the numerical setup, however, it is impossible to achieve exact $\Lambda_i = 0$ values, with bounds on the smallest computed Lyapunov exponents $\Lambda_\text{min} \sim 1/t$.

The evolution of the phase space vector $\vec\Psi$ is obtained from subsequent applications of the map $\hat U$. 
We use periodic boundary conditions $\psi_{N+1} = \psi_1$. The initial conditions for the amplitudes of the local complex components are drawn from an exponential distribution $p(x) = e^{-x}$, while their phases were generated as uncorrelated and random numbers chosen uniformly from the interval $[0, 2\pi ]$. The state vector is then uniformly rescaled such that the norm density $ \frac{1}{N}\sum|\psi_n|^2=1$. 
The largest integration time varied between $t_{\text{max}} = 10^8$ and $t_{\text{max}} = 10^9$. We have performed computations for a set of initial conditions to ensure the independence of results on the choice of initial state.
Unless stated otherwise the system size is set to $N = 200$. 

\begin{figure}
    \centering
    \includegraphics[width=\linewidth]{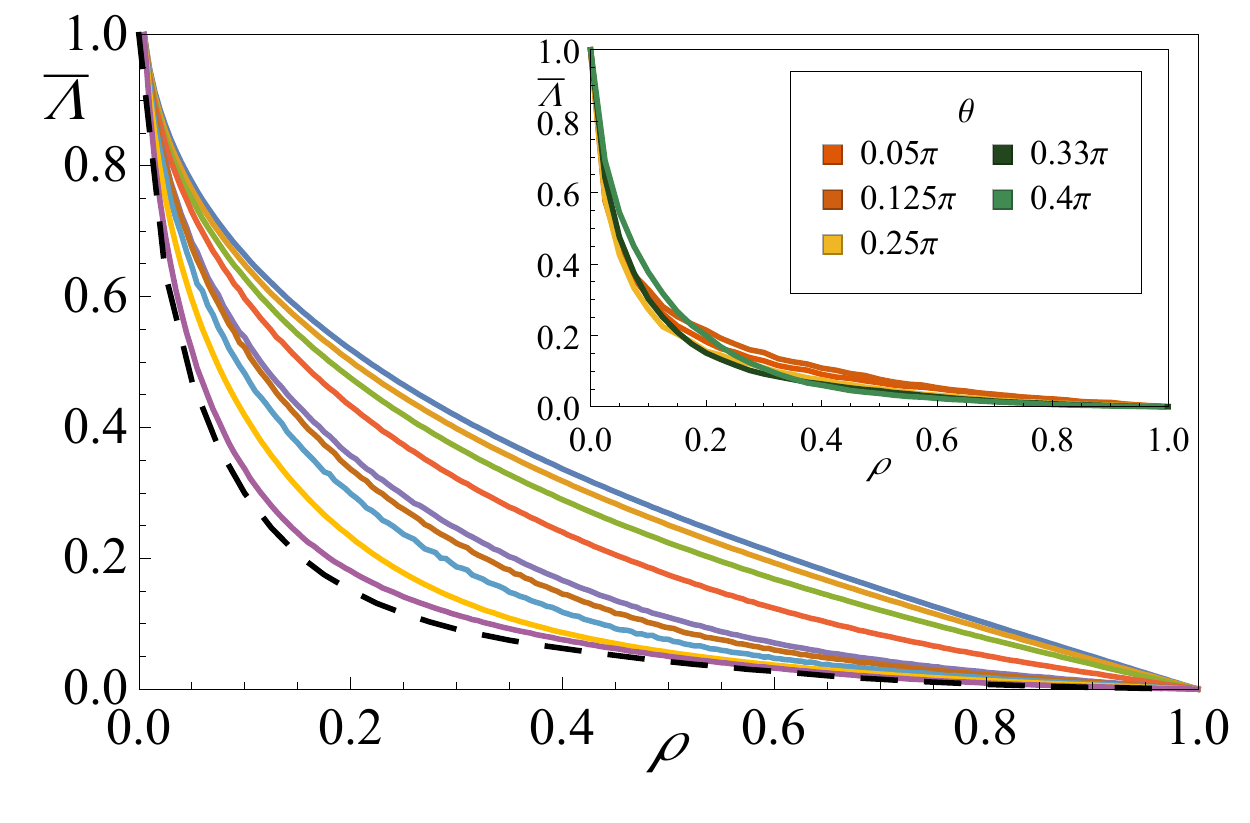}
    \caption{Renormalized Lyapunov spectrum $\lbrace \bar \Lambda_i \rbrace$ for LRN in proximity to the integrable limit. The angle $\theta = 0.33\pi$ is fixed. The deviation from integrable limit $g$ varies from $10^{-1}$ (blue) to $10^{-3}$ (purple). The dashed line is to show the asymptotic curve as $g \rightarrow 0$. In the inset we showcase the asymptotic curve as parameter $\theta$ is varied. For all cases system size $N = 200$.}
    \label{fig5}
\end{figure}

First, we show the dependence of the largest Lyapunov exponent $\Lambda_{max}$ on the distance $g$ or $\theta$ to the integrable limit in Fig. \ref{fig2} for both networks. Both curves show a dependence which might resemble a power law $\Lambda_{max} \sim g^\nu$ and $\Lambda_{max} \sim \theta^{\mu}$ with $\nu \approx 1/2$ and $\mu \approx 3/2$. Remarkably the SRN case shows a much slower diminishing of $\Lambda_{max}$ upon approaching the integrable limit as compared to the LRN. This is similar to the study of a Hamiltonian system dynamics \cite{danieli2019dynamical, mulansky2011strong, fine2013lyapunov}.
Our data in Fig.\ref{fig2} are obtained for two different system sizes $N=100,200$ and show very good agreement, therefore we can exclude
finite-size corrections.
We now proceed to the analysis of the entire Lyapunov spectrum. 
In Fig. \ref{fig3} and Fig. \ref{fig5} we show the renormalized Lyapunov spectrum $\bar \Lambda_i = \Lambda_i/\Lambda_{\text{max}}$ for SRN and LRN respectively. The index of Lyapunov exponents is rescaled $\rho= i/N$ so that all positive LCEs $\bar \Lambda(\rho)$ correspond to $\rho \in [0,1]$. We notice a dramatic qualitative difference between the two regimes. For the LRN case the renormalized Lyapunov spectrum  $\bar \Lambda(\rho)$ converges to a limiting smooth curve for $g\rightarrow 0$. For the SRN instead that curve vanishes in an exponential way. We will explain these observations in detail below.

For the SRN an increasing number of Lyapunov exponents seems to be vanishing upon approaching the integrable limit
as seen in Fig. \ref{fig3}. We replot the same spectrum in log scale in Fig. \ref{fig4} and notice an exponential decay of the renormalized spectrum: 

\begin{eqnarray}
\Lambda_\rho^\text{SRN} = \Lambda_\text{max} e^{-\rho/\beta} .
\label{finalSRN}
\end{eqnarray} 
We fit the exponential decay and plot the exponent $1/\beta$ versus $\theta$ in the inset in Fig. \ref{fig3}.
We observe that the exponent is rapidly diverging upon approaching the integrable limit such that $\beta(\theta\rightarrow 0) \rightarrow 0$.  
The entire Lyapunov spectrum of the SRN is therefore characterized by two scaling parameters - the largest Lyapunov exponent $\Lambda_{max}$ which is an inverse time scale, and the parameter $\beta$ which is an inverse length scale.  
This result explains and agrees with previous studies on dynamical glass in Hamiltonian systems \cite{danieli2019dynamical, mithun2019dynamical} where the largest Lyapunov exponent stems from local resonances with rapidly increasing distance between them upon approaching the integrable limit. Our results show that the Lyapunov spectrum contains the quantitative scaling parameters of that dynamical glass theory.

In contrast, the LRN spectrum is characterized by single parameter scaling. 
The renormalized Lyapunov spectrum approaches a smooth limiting curve $\bar \Lambda (\rho)$ as seen in Fig. \ref{fig5}. 
We compute the limiting curves by a linear fit of $\bar \Lambda_\rho(g)$ at each value of $\rho_j$.
Thus in the LRN regime, the final form of the spectrum is given by:
\begin{eqnarray}
\Lambda_\rho^\text{LRN} = \Lambda_\text{max}f(\rho, \theta)
\label{finalLRN}
\end{eqnarray}
The limiting curves for different values of $\theta$ are plotted in the inset of Fig.\ref{fig5} and show little if any variation. It appears
that the limiting curve $f(\rho)$ is universal for all LRN parameter choices. 

To further characterize the chaotic dynamics and showcase the difference between SR and LR networks we 
compute the Kolmogorov-Sinai entropy $K_\text{KS} = \int_0^1 \bar \Lambda_\rho d\rho$.
In the SRN case, from eq. \eqref{finalSRN} follows
\begin{eqnarray}
K_\text{KS}^\text{SRN} = \Lambda_\text{max}\beta(1-e^{-1/\beta})\;.
\end{eqnarray}
Therefore the renormalized Kolmogorov-Sinai entropy $k_\text{KS} = K_\text{KS}/\Lambda_{max}$  will
tend to zero in the integrable limit $k_\text{KS} \approx \beta$.

In the LRN regime the integral over the asymptotic function $f(\rho, \theta)$ (see eq. \eqref{finalLRN}) will lead to finite values of the renormalized KS entropy $k_\text{KS} = \int_0^1 f(\rho) d\rho > 0$ at the very integrable limit.

\textit{To conclude,} we identified the Lyapunov spectrum as a universal characteristic descriptor of the complex phase space dynamics of a macroscopic system in proximity to an integrable limit. The limit is characterized by a macroscopic number of conserved actions. We identify two classes of nonintegrable perturbation networks - short and long-range ones. Long-range networks are characterized by a single parameter scaling of the Lyapunov spectrum - knowing the
largest Lyapunov exponent allows to reconstruct the entire spectrum. Consequently all Lyapunov exponents scale as the largest one upon approaching the integrable limit. Typical long-range networks are realized with translationally invariant
lattice systems in the limit of weak nonlinearity. In that case, the actions correspond to normal modes extended over the entire real space. Nonintegrable perturbations will typically couple them all. On the other side, short-range networks
are characterized by a two-parameter scaling. In addition to the largest Lyapunov exponent, a diverging length scale
results in a suppression of the renormalized Lyapunov spectrum upon approaching the integrable limit. Typical short-range networks are realized with lattice systems and local (short-range) nonlinearities in the limit of weak coupling. 
Interestingly the short-range network case appears to include disordered systems as well. We, therefore, expect that
disordered systems with weak short-range nonlinearities will correspond to the SRN universality class. Quantizing the classical dynamics could lead to many-body localization in the case of short-range networks, as opposed to long-range networks.

\textit{Acknowledgments:}
This work was supported by the Institute for Basic Science (Project number: IBS-R024-D1).

\bibliography{bibliography}

\end{document}


\preprint{APS/123-QED}

\title{
Supplemental Material
}

\date{\today}

\maketitle

\section{Linear Problem}\label{supp:SectionLinearProblem}
In this section we provide equations of motion for the short range network. For convenience reasons we will use a bra-ket notation which slightly differs from the main text. Due to the nature of the evolution it is more convenient to represent the vector $\vec \Psi$ as a wave function constructed of $N/2$ unit cells with two components in each cell:
\begin{eqnarray}
        &&\ket{\Psi(t)} = \sum_{n = 1}^{N/2} \Big[\psi^A_n(t)\ket{A} + \psi^B_n(t)\ket{B}\Big] \otimes \ket{n} \nonumber \\ 
        &&= \sum_{n,\; p = A, B}\psi_n^p \ket{n, p}
\end{eqnarray}

We also rewrite the map $\hat U$ as an evolution operator using shift operation $\hat T$ and mixing operator $\hat C$:
\begin{equation}
    |\Psi(t+1)\rangle = \hat U^{(0)}|\Psi(t)\rangle=\hat T^{\dagger} \hat C \hat T \hat C |\Psi(t)\rangle
   \label{evol}
\end{equation}
Here the mixing operator $\hat C$ are $2\times 2$ is described by unitary matrices parametrized by the rotation angle $\theta$, which act locally on neighboring sites:
\begin{eqnarray}
    \hat{C}=\sum \hat c \otimes \ket{n}\bra{n} = \begin{pmatrix}
            \cos\theta & \sin\theta \\
            -\sin\theta & \cos\theta
            \end{pmatrix}\otimes|n\rangle\langle n| 
\end{eqnarray}
and $\hat T$ is the shifting operator moving all components of the lattice to the left:
\begin{eqnarray}
        \hat{T} = \sum_n \ket{A, n}\bra{B, n} + \ket{B, n}\bra{ A,n + 1}
\end{eqnarray}
The resulting equations of motion of the linear evolution are as follows: 
\begin{eqnarray}
&& \psi_n^{A}(t + 1) = \cos^2\theta\psi_n^A(t)- \cos\theta\sin\theta\psi_{n-1}^B(t)\nonumber  \\
&&+ \sin^2\theta\psi^A_{n +1}(t) + \cos\theta\sin\theta\psi_n^B(t) \nonumber \\
\nonumber\\
&& \psi_n^{B}(t + 1) = \sin^2\theta\psi_{n - 1}^B(t)- \cos\theta\sin\theta\psi_{n}^A(t) \nonumber  \\
&&+ \cos^2\theta\psi^B_{n}(t) + \cos\theta\sin\theta\psi_{n + 1}^A(t)
\label{eomsLinear}
\end{eqnarray}

\begin{figure}
    \centering
    \includegraphics[width = 0.45\textwidth]{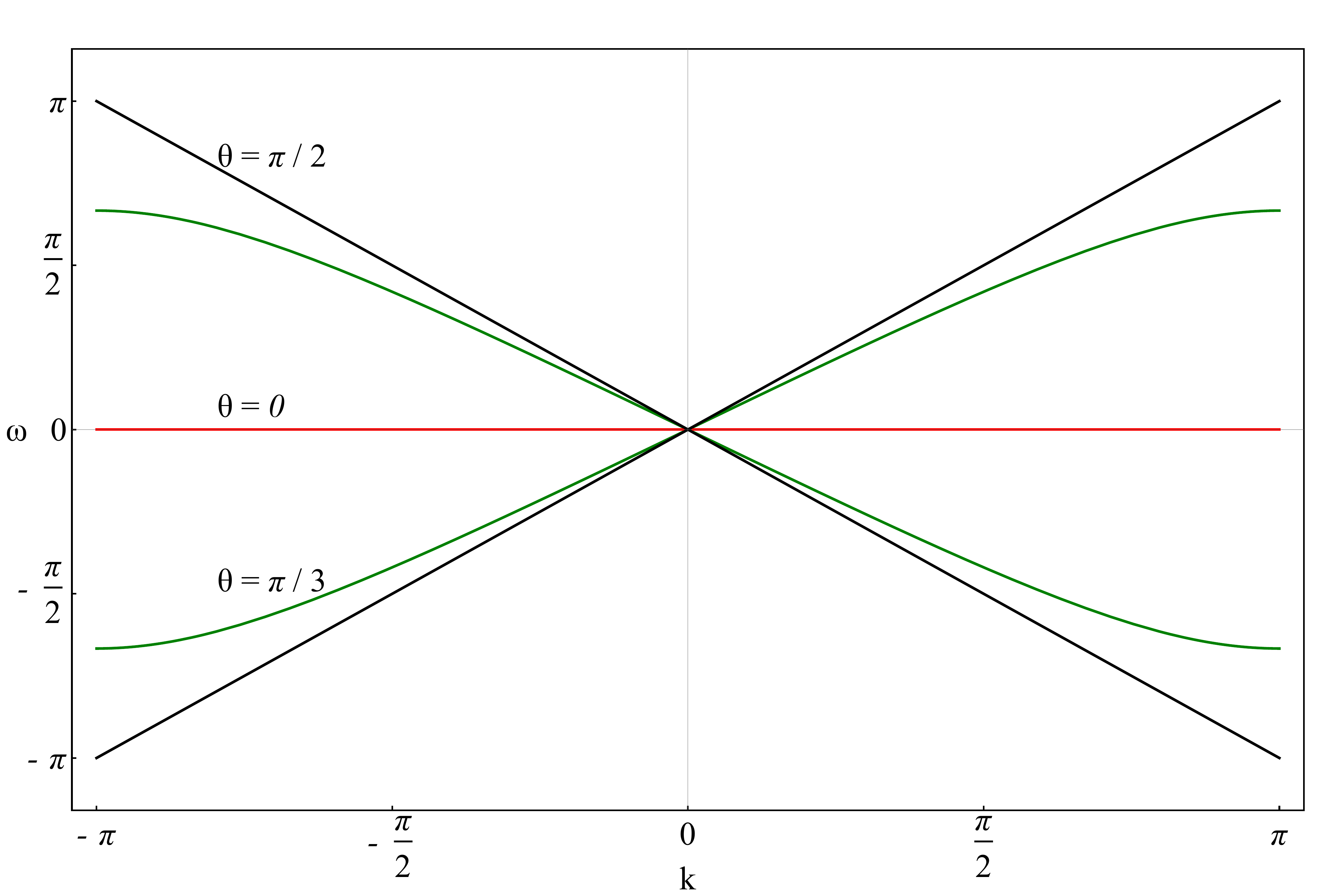}
    \caption{The dispersion relation corresponding to Unitary Circuits (see eq. \eqref{supp:dispersion}). The parameter $\theta$ is varied to showcase a dispersionless flat band (red), a case of constant group velocity case (black) and the generic case (green).}
    \label{fig:spectrum}
\end{figure}

The solution comes in terms of plain waves $\vec{\psi}_n = e^{i k n}\vec{\psi}_k$, where $\vec{\psi}$ is a two component vector and $k$ is a wave number. The final diagonalization procedure leads to the dispersion relation (see Fig. \ref{fig:spectrum})
\begin{equation}
\cos\omega = \cos^2\theta + \sin^2\theta\cos k
\label{supp:dispersion}
\end{equation}

\section{Short Range Network}\label{supp:SectionSRN}
For convenience we rewrite the linear part of the evolution as follows:
\begin{eqnarray}
&& \psi_n^{A}(t + 1) = \alpha^A_n(\Psi(t)) \nonumber \\
&& \psi_n^{B}(t + 1) = \alpha^B_n(\Psi(t))
\end{eqnarray}

The nonlinearity inducing operator $\hat G$ is applied after the linear part of the evolution:
\begin{eqnarray}
\hat U_\text{nonlin} = \hat  G \hat U^{(0)}
\label{nonlinEvol}
\end{eqnarray}

The nonlinearity is induced through an additional norm-dependent phase rotation depending on the result of the local linear evolution:
\begin{eqnarray}
\hat G = \sum_{n, p} e^{i g |\alpha^p_n|^2}\otimes\ket{n , p}\bra{n, p}
\end{eqnarray}

The final equations of motion are:
\begin{eqnarray}
&& \psi_n^{A}(t + 1) = e^{i g |\alpha^A_n|^2}\big[\cos^2\theta\psi_n^A(t)- \cos\theta\sin\theta\psi_{n-1}^B(t)\nonumber  \\
&&+ \sin^2\theta\psi^A_{n +1}(t) + \cos\theta\sin\theta\psi_n^B(t)\big] \nonumber \\
\nonumber\\
&& \psi_n^{B}(t + 1) = e^{i g |\alpha^B_n|^2}\big[\sin^2\theta\psi_{n - 1}^B(t)- \cos\theta\sin\theta\psi_{n}^A(t) \nonumber  \\
&&+ \cos^2\theta\psi^B_{n}(t) + \cos\theta\sin\theta\psi_{n + 1}^A(t)\big]
\label{eomsNonLinear}
\end{eqnarray}
The equations of motion are next to nearest neighbor form which falls under the definition of a short range network. The integrable limit is reached for $\theta = 0$. The system turns integrable and the equations of motion read:
\begin{eqnarray}
&& \psi_n^{A}(t + 1) = e^{i g |\psi^A_n|^2}\psi_n^A(t) \nonumber \\
&& \psi_n^{B}(t + 1) = e^{i g |\psi^B_n|^2}\psi_n^B(t)
\label{eomsSRNintegrable}
\end{eqnarray}

\section{Long range network}\label{supp:SectionLRN}

The long range network of observables is obtained in the normal mode space of the model. The wave function can be represented as a sum of normal modes:
\begin{equation}
    \ket{\Psi(t)} = \sum_{k, r} e^{i \omega_k^r t} c_k^r(t) \ket{\psi_k^r},
    \label{wfNormalModeRepr}
\end{equation}
where the index $r$ corresponds to one of the two bands and $\ket{\psi_k^r}$ is a corresponding normal mode. In the linear setup the normal mode coefficients are conserved in time $c_k^r(t) = const$ and as such are integrals of motion. In the reciprocal space the network consists of disconnected nodes with $c_k^r$ associated to each node. 
Let us expand the nonlinear evolution operator eq.\eqref{nonlinEvol} for small values of the parameter $g$: 
\begin{eqnarray}
\hat U_\text{nonlin} = \hat U^{(0)} +i g \sum_{n,p} |\alpha_n^p|^2 \hat U^{(0)},
\label{nonlinEvolExpanded}
\end{eqnarray}
where $|\alpha_n^p|^2$ can be represented as:
\begin{eqnarray}
A_n^p = \bra{\Psi(t)}\hat U^{0}\ket{n, p}\bra{n, p}\hat U^{0}\ket{\Psi(t)}.
\end{eqnarray}
Using the normal mode representation of the wave function \eqref{wfNormalModeRepr} we obtain the evolution equations of normal mode coefficients: 
\begin{eqnarray}
   &&  c_k^r(t+1) = e^{i \omega^r_k} c_k^r(t) + \nonumber \\ 
   && ig\sum_{r_1,r_2,r_3}\sum_{k_1,k_2,k_3} I_{k,k_1,k_2,k_3}^{r,r_1,r_2,r_3}c^{r_1}_{k_1}(t)c^{r_2}_{k_2}(t)\left(c^{r_3}_{k_3}(t)\right)^* ,
   \label{EoMs_EOs}
\end{eqnarray}
where the overlap integrals $I$ are given by the following expression:
\begin{eqnarray}
   &&I_{k,k_1,k_2,k_3}^{r,r_1,r_2,r_3} = e^{i \omega_{k_1}}\sum_{n,p}\bra{n,p}\hat U^{(0)}\ket{\psi_{k_2}^{r_2}}\bra{\psi_{k_3}^{r_3}}\hat U^{(0)}\ket{n, p} \nonumber \\
   &&= e^{i (\omega_{k_1}^{r^1} + \omega_{k_2}^{r^2} - \omega_{k_3}^{r^3})}\sum_{n,p}\langle n,p \ket{\psi_{k_2}^{r_2}}\langle{\psi_{k_3}^{r_3}}\ket{n, p}.
\end{eqnarray}
The number of triplet terms induced by nonlinearity in equation \eqref{EoMs_EOs} is proportional to $N^3$. These equations correspond to a long range network.

\section{Deviation vectors}\label{supp:SectionDeviationVectors}
\begin{figure}
\centering
\begin{subfigure}{\linewidth}
   \includegraphics[width=\linewidth]{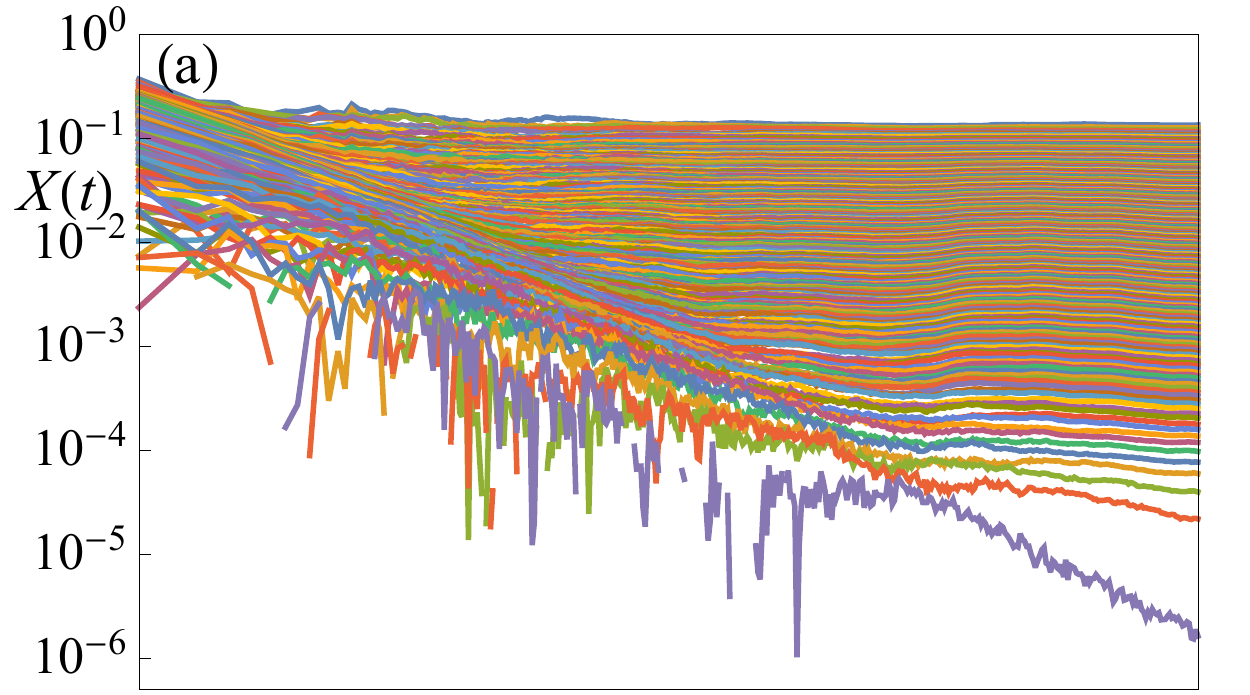}
\end{subfigure}
\begin{subfigure}{\linewidth}
   \includegraphics[width=\linewidth]{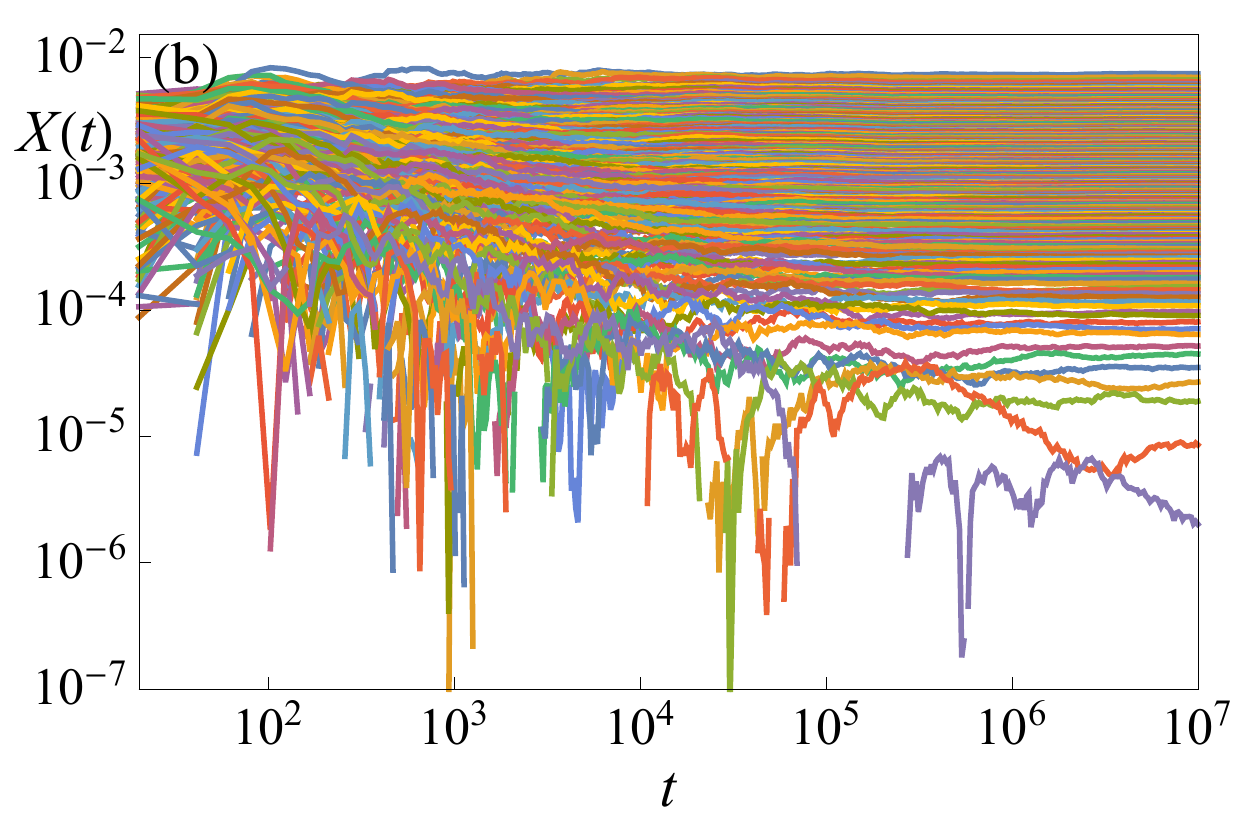}
\end{subfigure}
\caption{The evolution of positive transient Lyapunov exponents. a) SRN case with angle $\theta = 0.1$ and nonlinearity $g = 1.0$, b) LRN case with angle $\theta = 0.33\pi$ and $g = 0.1$. For both cases system size $N = 200$. }
\label{supp:fig2}
\end{figure}

To compute the set of Lyapunov exponents we follow the evolution of tangent vectors $\lbrace \vec w_i \rbrace$. Each vector corresponds to the direction of the exponential growth or shrinking of the deviation from the initial trajectory - in total $2 N$ vectors. The evolution of tangent vectors is done using the corresponding equations of motion derived below. We measure the magnitude of growth  $\gamma(t) = |\vec w(t)|$ of each tangent vector and compute transient Lyapunov exponents $X_i(t) = 1/t\sum_\tau^t \log \gamma(\tau)$ after which the tangent vectors are orthonormalized using a Gram-Schmidt procedure. The evolution of positive transient Lyapunov exponents $X(t)$ is shown in Fig. \ref{supp:fig2}. After an initial decay the transient Lyapunov exponents saturate. The saturated values are taken as final values for Lyapunov exponents $\Lambda$. Due to the conservation of the norm two exponents are expected to attain zero value. In the figure we see one of them (bottom most purple line) tending to zero with increasing time and no saturation. 

\subsection{Equations of Motion}
We start from the nonlinear EoM \eqref{eomsNonLinear}:
\begin{eqnarray}
&& \psi_n^{A}(t + 1) = e^{ i g |\alpha_n^A(\Psi(t))|^2} \alpha_n^A(\Psi(t)) \nonumber \\
&& \psi_n^{B}(t + 1) = e^{ i g |\alpha_n^B(\Psi(t))|^2} \alpha_n^B(\Psi(t)),
\end{eqnarray}
where $\alpha^{A,B}_n$ are linear functions of the local components of the wave function $\ket{\Psi(t)}$ according to equations $\eqref{eomsLinear}$. We consider a small deviation $\vec{\varepsilon}(t)$ from the initial trajectory $\vec{x}(t)$:
\begin{eqnarray}
\vec{\psi} = \vec{x} + \vec{\varepsilon}
\end{eqnarray}
Substituting into \eqref{eomsNonLinear}:
\begin{eqnarray}
&& \psi_n^{A}(t + 1) = e^{ i g |\alpha^A_n[\vec{x}(t)+\vec{\varepsilon}(t)]|^2} \alpha^A_n\left[(\vec{x}(t)+\vec{\varepsilon}(t))\right] \nonumber \\
&& \psi_n^{B}(t + 1) = e^{ i g |\alpha^B_n[\vec{x}(t)+\vec{\varepsilon}(t))]|^2} \alpha^B_n\left[(\vec{x}(t)+\vec{\varepsilon}(t))\right].
\label{nonlinEomInsertedSols}
\end{eqnarray}
Expanding the nonlinear term and keeping terms only in the $1$st order of $\vec{\varepsilon}$ results in
\begin{eqnarray}
   && |\alpha^p_n[\vec{x}(t)+\vec{\varepsilon}(t)]|^2 = |\alpha^p_n[\vec{x}(t)]+\alpha^p_n[\vec{\varepsilon}(t)]|^2 = \nonumber \\ &&\alpha^p_n(\vec{x}(t))[\alpha^p_n(\vec{x}(t))]^* + \alpha^p_n(\vec{\varepsilon}(t))[\alpha^p_n(\vec{\varepsilon}(t))]^* + \nonumber \\ &&\alpha^p_n(\vec{\varepsilon}(t))[\alpha^p_n(\vec{x}(t))]^* + \alpha^p_n(\vec{x}(t))[\alpha^p_n(\vec{\varepsilon}(t))]^* \approx  \nonumber \\
   && |\alpha^p_n(\vec{x}(t))|^2 + \Delta^p_n(t),
\end{eqnarray}
where
\begin{eqnarray}
   &&\Delta^p_n(t) = \alpha_n^p(\vec{x}(t))[\alpha_n^p(\vec{\varepsilon}(t))]^* + c.c. \nonumber \\
\end{eqnarray}
Thus we can rewrite the exponential term by expanding $e^{i g \Delta_n^{A,B}(t)}$:
\begin{eqnarray}
   e^{ i g |\alpha_n^p[\vec{x}(t)+\vec{\varepsilon}(t))]|^2} = e^{ i g |\alpha_n^p(\vec{x}(t))|^2}\left[1 + i g \Delta_n^p(t)\right] 
\end{eqnarray}
With \eqref{nonlinEomInsertedSols} and using the linearity of $\alpha_n^p$ we finally arrive at the
following linear equations:
\begin{eqnarray}
   &&\varepsilon_n^p(t+1) = e^{ i g |\alpha_n^p(\vec{x}(t))|^2}\Big\lbrace \alpha_n^p[\vec{\varepsilon}(t)] + i g \Delta_n^p(t)\alpha_n^p[\vec{x}(t)]\Big\rbrace. \nonumber \\
\end{eqnarray}